\documentstyle[12pt]{article}
\begin{document}
\thispagestyle{empty}
\begin{center}
\LARGE \tt \bf{Gravitational torsion kinks and thick domain walls.}
\end{center}
\vspace{2.5cm}
\begin{center} {\large L.C. Garcia de Andrade\footnote{Departamento de F\'{\i}sica Teorica - UERJ.
Rua S\~{a}o Fco. Xavier 524, Rio de Janeiro, RJ
Maracan\~{a}, CEP:20550-003 , Brasil.
E-Mail.: GARCIA@SYMBCOMP.UERJ.BR}}
\end{center}
\vspace{2.0cm}
\begin{abstract}
The dynamics of a gravitational torsion kink as a plane symmetric thick 
domain wall solution of Einstein-Cartan (EC) field equation is given.
The spin-torsion energy has to be as high as the gravitational kink 
potential otherwise torsion will not contribute as an appreciable effect to 
domain wall.Cartan torsion also contributes to the orthonal pressure of the domain wall.
\end{abstract}
\vspace{2.0cm}
\newpage
Riemannian \cite{1,2} and non-Riemannian \cite{3,4} topological defects have
been applied with success in many areas of modern physics including 
cosmology \cite{5},Helium superfluid \cite{6,7} and high-energy physics 
\cite{8} where a QCD string from a quark crosses an axionic domain wall.
Other examples of domain walls have been found in the literature \cite{6}.
Torsion strings have been extensively investigated by Letelier \cite{9,10} 
and myself \cite{11,12,13}.Besides some application of torsion strings to 
superfluids, where a domain wall crossed by string can be physically 
interpreted as a vortice,has been investigated by Volovik \cite{7}.The geometry of thin planar walls with torsion have been recently investigated by myself \cite{11,12}.However a more detailed of the dynamics of thick domain wall was missing to complete this study, the only close attempt to the problem of torsion kinks have been addressed by Baekler et al.\cite{14} for a massless Higgs field which of course differs from the treatment and purposes we have here.In this letter asimple solution of EC field equations where Cartan torsion is given and compared with the gravitational kink potential have been considered.Let us consider the most general plane symmetric metric
\begin{equation}
ds^{2}=e^{{\nu}(z,t)}dt^{2}-e^{{\lambda}(z,t)}dz^{2}-e^{{\psi}(z,t)}(dx^{2} + dy^{2})
\label{1}
\end{equation}
The energy-stress tensor is composed of two parts, first the scalar field 
$ {\Phi} $ with self-interactions contained in a potential $V({\Phi})$ 
given by 
\begin{equation}
{T_{ij}}^{Kink}= {\partial}_{i}{\Phi}{\partial}_{j}{\Phi}-g_{ij}(\frac{1}{2}{\partial}_{k}{\Phi}{\partial}^{k}{\Phi}-V({\Phi}))
\label{2}
\end{equation}
the other part is the torsion energy stress tensor given by
\begin{equation}
{{T_{i}}^{k}}^{torsion}={S_{iml}S^{kml}-\frac{1}{2}{{\delta}_{i}}^{k}S^{2}}
\label{3}
\end{equation}
where we have considered that the spin density tensor is totally skew symmetric as in the case of Dirac electrons and where $ S_{ijk} $ is the torsion tensor and $S^{2}$ is the torsion-spin energy.The scalar field depends only on the z-coordinate which is orthogonal to the wall,i.e,we have ${\Phi}(z)$.The EC equation can be writen in the quasi-Einsteinian form
\begin{equation}
{G_{ij}}^{Riem}=kT_{ij}
\label{4}
\end{equation}
where $ T_{ij} $ is the sum of the stress-energy tensors given by {\ref{2}} 
and {\ref{3}} and ${G_{ij}}^{Riem}$ is the Riemannian-Einstein tensor.The 
components of the stress-energy tensors are given by
\begin{equation}
{T^{t}}_{y}={T^{x}}_{x}={T^{y}}_{y}=e^{-{\lambda}}{{\Phi}'}^{2}+V({\Phi})-S^{2}={\rho}_{eff} 
\label{5}
\end{equation}
and 
\begin{equation}
{{T}_{z}}^{z}=-\frac{1}{2}(e^{-{\lambda}}{{\Phi}'}^{2}+V({\Phi})-S^{2})= -p_{eff}
\label{6}
\end{equation}
where $ {\rho}_{eff} $ and $ p_{eff} $ are respectively the effective 
density and pressure containing torsion energy and the potential 
gravitational kink.The scalar field equations are obtained by using the 
minimal coupling between torsion and the derivative of the scalar fields 
since it is well known that torsion does not coupled directly with the 
scalar fields.This situation is very similar to what happens in the 
Schr\"{o}dinger equation \cite{15} in space-times with torsion.The scalar 
field equation with torsion is
\begin{equation}
{{\Phi}"}+{{\Phi}'}({\psi}'-S)=-\frac{dV}{d{\Phi}}e^{\lambda} 
\label{7}
\end{equation}
where the last term on the LHS represents the interaction between torsion 
component $ S(z) $ and the derivative of the scalar field with respect to 
the z-coordinate.To simplify matters we consider solutions where the 
following constraint is obeyed
\begin{equation}
{\psi}=S(z)
\label{8}
\end{equation}
Expression {\ref{8}} helps to simplify equation {\ref{7}} to
\begin{equation}
{{\Phi}"}=-\frac{dV}{d{\Phi}}e^{\lambda} 
\label{9}
\end{equation}
As shown by Goetz \cite{16} in his Riemannian version of the problem 
considered here,the metric can be simplified to 
\begin{equation}
ds^{2}=A(z)(dt^{2}-dz^{2}-b(t)(dx^{2}+dy^{2}))
\label{10}
\end{equation}
Note that the time dependence of $ g_{tt} $ has been eliminated.The EC 
field equation becomes
\begin{equation}
{G^{t}}_{t}-{G^{x}}_{x}=0
\label{11}  
\end{equation}
which implies the following equation
\begin{equation}
{b}^{..}b-{{b}^{.}}^{2}=0
\label{l2}  
\end{equation}
The immediate solution of this equation is
\begin{equation}
b(t)=e^{ct}
\label{13}
\end{equation}
The other equations are 
\begin{equation}
-\frac{{{A}"}}{A^{2}}+\frac{3}{2}{\frac{{{A}'}^{2}}{A^{2}}}-\frac{c^{2}}{2A}=\frac{8{\pi}G{{\Phi}'}^{2}}{A}
\label{14}
\end{equation}
and 
\begin{equation}
-\frac{{{A}'}^{2}}{A^{2}}+\frac{c^{2}}{A}=16{\pi}GV({\Phi})-2S^{2} 
\label{15}
\end{equation}
Since $ {\psi}'=\frac{{A}'}{A}=S $ equations {\ref{13}} and {\ref{14}} 
simplify to
\begin{equation}
A=\frac{c^{2}}{16{\pi}GV({\Phi})-S^{2}}
\label{16}
\end{equation}
Thus expression {\ref{16}} represents a solution for the potential 
$ V({\Phi})$.The kink torsion is obtained for the potential
\begin{equation}
V({\Phi})=V_{0}{cos}^{2(1-n)}\frac{\Phi}{f}
\label{17}
\end{equation}
The other field equation yields the torsion energy 
\begin{equation}
S^{2}=\frac{c^{2}+16{\pi}G{{\Phi}'}^{2}}{A(z)}
\label{18}
\end{equation}
The final metric is given by
\begin{equation}
ds^{2} = \frac{c^{2}}{16{\pi}V({\Phi})-S^{2}}[dt^{2}-dz^{2}-e^{ct}(dx^{2}+dy^{2})) 
\label{19}
\end{equation}
Note that if the torsion energy is much weaker than the scalar kink 
potential the metric (\ref{19}) does not depend on torsion.Therefore as 
usual in torsion physics it is necessary the torsion be considered in a 
high energy context like in the core of neutron stars.When the potential 
$V({\Phi})$ coincides with the torsion energy the metric explodes as is 
usual in the locus of the walls.
\section*{Acknowledgements}
I am very much indebt to Professors P.S.Letelier,A.Wang,I.Shapiro and E.Mielke for helpful discussions on the subject of this paper.Thanks are also due to CNPq. (Brazilian Government Agency) and to UERJ for financial support.Special thanks go to Prof.G.E.Volovik for sending me many of his repreints dealing with torsion strings and domain walls in superfluids.

\end{document}